\documentclass[aps,superscriptaddress,showkeys,showpacs,floatfix,amsmath,amssymb,twocolumn]{revtex4}
\usepackage{graphicx}
\usepackage{epsfig}
\usepackage{dcolumn}
\usepackage{bm}

\renewcommand{\r}{{\bf r}}

\newcommand{\be}{\begin{eqnarray}}
\newcommand{\ee}{\end{eqnarray}}

\begin{document}
\title{Odd-even mass difference and isospin dependent pairing interaction}

\author{C. A. Bertulani}
\affiliation{Department of Physics, Texas A\&M University-Commerce,
Commerce, Texas 75429, USA}

\author{H. F. L\"u }
\affiliation{College of Science, China Agricultural University,
Beijing, P.R.China}


\author{H. Sagawa}
\affiliation{Center for Mathematics and Physics, University of Aizu,
Aizu-Wakamatsu, 965-8580 Fukushima, Japan}\date{\today}

\begin{abstract}
The neutron and proton  odd-even mass differences are  studied with
Hartree-Fock+BCS (HFBCS) calculations with Skyrme interactions  and
an isospin dependent contact pairing interaction, which is recently
derived from a microscopic nucleon-nucleon interaction. To this end,
we perform HFBCS calculations for even and odd semi-magic Tin and
Lead isotopes together with even and odd Z isotones with  $N$= 50
and 82. The filling approximation is applied to  the last unoccupied
particle in odd nuclei.   Comparisons with the experimental data
show a clear manifestation of the isospin dependent pairing
correlations in both proton and neutron pairing gaps.

\end{abstract}


\maketitle

It has been  known that pairing correlations
 play an important role in finite and also infinite
nuclear systems \cite{boh58,BM69,BB05}.
   Recently, the theory of nuclear masses
 or binding
energies has attracted renewed interest with the advent of
self-consistent mean field theories, and also density functional
theories (DFT) \cite{Ben03,Stoi06}.  A global  feature of the nuclear
 binding energies is the odd-even mass staggering (OES) phenomenon.
   Several theoretical studies have been made to attribute this phenomenon
 to the BCS superfluidity in the nuclear ground states.
  It has been  pointed out that other effects also
  contribute the OES effect \cite{Satula98,dug01}. 

  Recently, global calculations of nuclear masses became feasible
  by using modern computational resources.  A goal of these global calculations
 is to improve the reliability of theories and to establish  universal
  energy density functionals for nuclear masses.
  In this respect, the pairing correlations should be carefully examined by using
  microscopic methods such as Hartree-Fock(HF)+BCS or Hartree-Fock-Bogoliubov
 (HFB) theories.  Indeed, first studies in this direction have been carried out and  a
 possible isospin dependence of the effective pairing
 interaction has been discussed in the literature \cite{mar07,Yama08}.

   The nuclear interaction may conserve the isospin at a fundamental level, but
  core polarization can induce isospin dependence when the core has
 a neutron excess.  Another contribution may come from the Coulomb interaction.
  Recently, an effective isospin dependent pairing interaction was proposed
  from the study of nuclear matter pairing gaps calculated by realistic
  nucleon-nucleon interactions.
  In ref. \cite{mar07},
the density$-$dependent pairing interaction was defined by \be
V_{pair}(1,2)= \mathrm{V}_0 \,\mathrm{g}_\tau[\rho,\beta\tau_z]
\,\delta(\r_1-\r_2),
\label{eq:pairing_interaction} \ee
where $\rho=\rho_n+\rho_p$ is the
nuclear density and $\beta$ is the asymmetry parameter
 $\beta=(\rho_n -\rho_p)/\rho$.
The isovector dependence is introduced through  the
density-dependent term $\mathrm{g}_\tau$. The function
$\mathrm{g}_\tau$ is determined by the  pairing gaps in nuclear
matter and its functional form is given by
\be
\mathrm{g}_\tau[\rho,\beta\tau_z] =  1
-\mathrm{f}_\mathrm{s}(\beta\tau_z)\eta_\mathrm{s}
\left(\frac{\rho}{\rho_0}\right)^{\alpha_\mathrm{s}}
-\mathrm{f}_\mathrm{n}(\beta\tau_z)\eta_\mathrm{n}
\left(\frac{\rho}{\rho_0}\right)^{\alpha_\mathrm{n}} \;,
\label{eq:g1t}
\ee where $\rho_0$=0.16~fm$^{-3}$ is the saturation
density of symmetric nuclear matter. We choose
$\mathrm{f}_\mathrm{s}(\beta\tau_z)=1-\mathrm{f}_\mathrm{n}(\beta\tau_z)$
and
$\mathrm{f}_\mathrm{n}(\beta\tau_z)=\beta\tau_z=\left[\rho_\mathrm{n}({\bf r})-\rho_\mathrm{p}({\bf r})
\right]\tau_z/\rho({\bf r})$. The parameters for $g_\tau$ are obtained from the fit to the pairing gaps in
symmetric and neutron matter obtained by the microscopic
nucleon-nucleon interaction. The pairing strength $V_0$ will be
  adjusted to give the best fit
to odd-even staggering of nuclear masses.

\begin{table}[h]
\begin{center}
\setlength{\tabcolsep}{.06in}
\renewcommand{\arraystretch}{1.5}
\begin{tabular}{cccccccc}
\toprule
interaction & $V_0$ (MeV)& $\rho_0$ fm$^{-3}$ & $\eta_\mathrm{s}$ & $\alpha_\mathrm{s}$ & $\eta_\mathrm{n}$ & $\alpha_\mathrm{n}$ \\
\colrule
$g_\tau$ & 824  & 0.16 fm & 0.677& 0.365 & 0.931  & 0.378  \\
$g_s$ & 1400  & 0.16 fm  & 1. & 1. & ---  & ---  \\
\botrule  
\end{tabular}
\end{center}
\caption{Parameters for the density-dependent function
$\mathrm{g}_\tau$ defined in Eq.~(\ref{eq:pairing_interaction}) (first row)
and $g_s$ in Eq. \eqref{isoscalar}.
The parameters for $g_\tau$ are obtained from the fit to the pairing gaps in
symmetric and neutron matter obtained by the microscopic
nucleon-nucleon interaction. The paring strength $V_0$
  is adjusted to give the best fit
to odd-even staggering of nuclear masses. The parameters for $g_s$
correspond to a surface peaked pairing interaction with no isospin
dependence. The parameters in this case are  adjusted to a best
global fit of nuclear masses \cite{bertsch09}.}  
\end{table}%

In the the original EV8 code \cite{EV8}, a pure contact interaction
was used without the isospin dependence. In our notation, this amounts replacing
the isospin dependent function $g_\tau$ in Eq. \eqref{eq:pairing_interaction}  by the isoscalar
function
 \begin{equation}
g_s=1-\eta_\mathrm{s}
\left(\frac{\rho}{\rho_0}\right)^{\alpha_\mathrm{s}}
\label{isoscalar}.
\end{equation}
 The parameters in this
case were adjusted to a best global fit of nuclear masses \cite{bertsch09}. They correspond to
a surface peaked pairing interaction. Table I shows the parameters for 
$g_\tau$ and $g_s$ used in the present work.

In several previous publications \cite{dug01,mar07,Yama08},
 the OES was not obtained from the differences of calculated
binding energies, but rather inferred from the average HFB gap parameters.
  It should be mentioned that the average HFB gaps are sometimes
 substantially different from the calculated odd-even mass differences.
 In this work,  we compare directly the calculated OES with the experimental
  ones.  There are several prescriptions to obtain the OES such as
 3-point, 4-point, and 5-point formulas.  We adopt the 3-point formula
  $\Delta^{(3)}$  centered at odd nucleus, i.e., odd-N nucleus for neutron gap and
 odd-Z nucleus for proton gap \cite{BM69}:
\be
\Delta^{(3)}(N,Z)\equiv-\frac{\pi_{N}}{2} \Big[\mathrm{B}(N-1,Z)&-&2\mathrm{B}(N,Z)   \\\nonumber
&+&\mathrm{B}(N+1,Z)\Big] \; , \label{eq:oes} \ee where $B(N,Z)$ is
the binding energy of (N,Z) nucleus and
 $\pi_N=(-)^N$ is the number parity.
For even nuclei, the OES is known to be sensitive not only to the
pairing gap, but also to mean field effects, i.e., shell effects
and deformations~\cite{Satula98,dug01}.  Therefore, the comparison of a
theoretical pairing gap with OES should be done with caution.
  One advantage of $\Delta^{(3)}_o$ ($N=$ odd in eq. \eqref{eq:oes})
is the suppression of the contributions from the mean field to the
gap. At a shell closure, the OES~(\ref{eq:oes}) does not go to zero
as expected, but it increases substantially. This large gap is an
artifact due to the shell effect, which is totally independent of
the pairing gap itself.

 We use the code EV8 ~\cite{EV8} to carry out the HF+BCS calculations
 with Skyrme interactions.  The pairing interaction  \eqref{eq:g1t} adopted
 is a contact interaction and can be used in a properly
  truncated configuration space.
 In the present study, the energy window is taken as 10 MeV around the
Fermi level as is ref. \cite{EV8}.
This is a limitation of the EV8 code, which solves the
HF+BCS equations via a discretization of the individual
wavefunctions on a three-dimensional Cartesian mesh,
   while this program allows a
flexibility in the determination of the nuclear shape.
For a global study of OES, it is important to allow the flexibility of
triaxial shapes.

First, the HF+BCS calculations are performed for even-even nuclei.
The variables in the theory are the orbital wave functions $\phi_i$
and the BCS amplitudes $v_i$ and $u_i = \sqrt{1 - v^2_i}$. By
solving the BCS equations for the amplitudes, one obtains the
pairing energy from
\begin{equation}
E_{pair} = \sum_{i\neq j} V_{ij} u_i v_i u_j
v_j + \sum_i V_{ii} v^2_i \label{Epair}
\end{equation}
where $V_{ij}$ are the matrix elements
of the pairing interaction, Eq. \eqref{eq:pairing_interaction}, namely
$$
V_{ij}=V_0\int d^3 r |\phi_i({\bf r})|^2 |\phi_j({\bf r})|^2
\mathrm{g}_\tau[\rho ({\bf r}),\beta({\bf r})\tau_z],
$$
where $\rho({\bf r})=\sum_i v_i^2 |\phi_i({\bf r})|^2$.

 In the present study, we take sub-closed shell nuclei only so
that the HF minimum appears essentially around the spherical
configurations. After determining the single-particle energies of
even-even nuclei,
 the odd-A nuclei are calculated with the so-called filling
approximation for the odd particle starting from the HF+BCS
solutions of neighboring even-even nuclei: ones selects an orbital
$i$ to be blocked, and changes the BCS parameters $v^2_i$ and $u_i
v_i$ for that orbital. The change is to set $v^2_i = u^2_i=1/2$  in
Eq. \eqref{Epair} for the pairing energy at an orbital near the Fermi
energy. Note that the filling approximation gives equal occupation
numbers to both time-reversed partners, and does not account for the
effects of time-odd fields. More details of the procedure are
presented in ref. \cite{bertsch09}.


The HF+BCS calculations are performed by using SLy4 
and SkP Skyrme interactions.  The iteration procedure used in EV8
achieves an accuracy of about 100 keV, or less, in 500 iterations.
For the pairing channels we take the surface-type contact
interaction, Eq. \eqref{isoscalar}, and the isospin dependent
interaction, Eq. \eqref{eq:g1t}.  The density dependence of the
latter one is essentially the mixed-type interaction between the
surface and the volume types. The pairing strength $V_0$ depends on
the energy window adopted for BCS calculations. The odd nucleus is
treated in the filling approximation, by blocking one of the
orbitals. The blocking candidates are chosen within an energy window
of 10 MeV around the Fermi energy.
This energy window is rather small, but it is the maximum allowed by 
the program EV8.  It is shown that the  EV8 model gives 
almost  equivalent results 
 to the HF+Bogoliubov model with a larger energy window,
 except unstable nuclei very close to the neutron drip line \cite{bertsch09}. 

\begin{figure}[htb]
\vspace{-0.5cm}
\includegraphics[clip,scale=0.375]{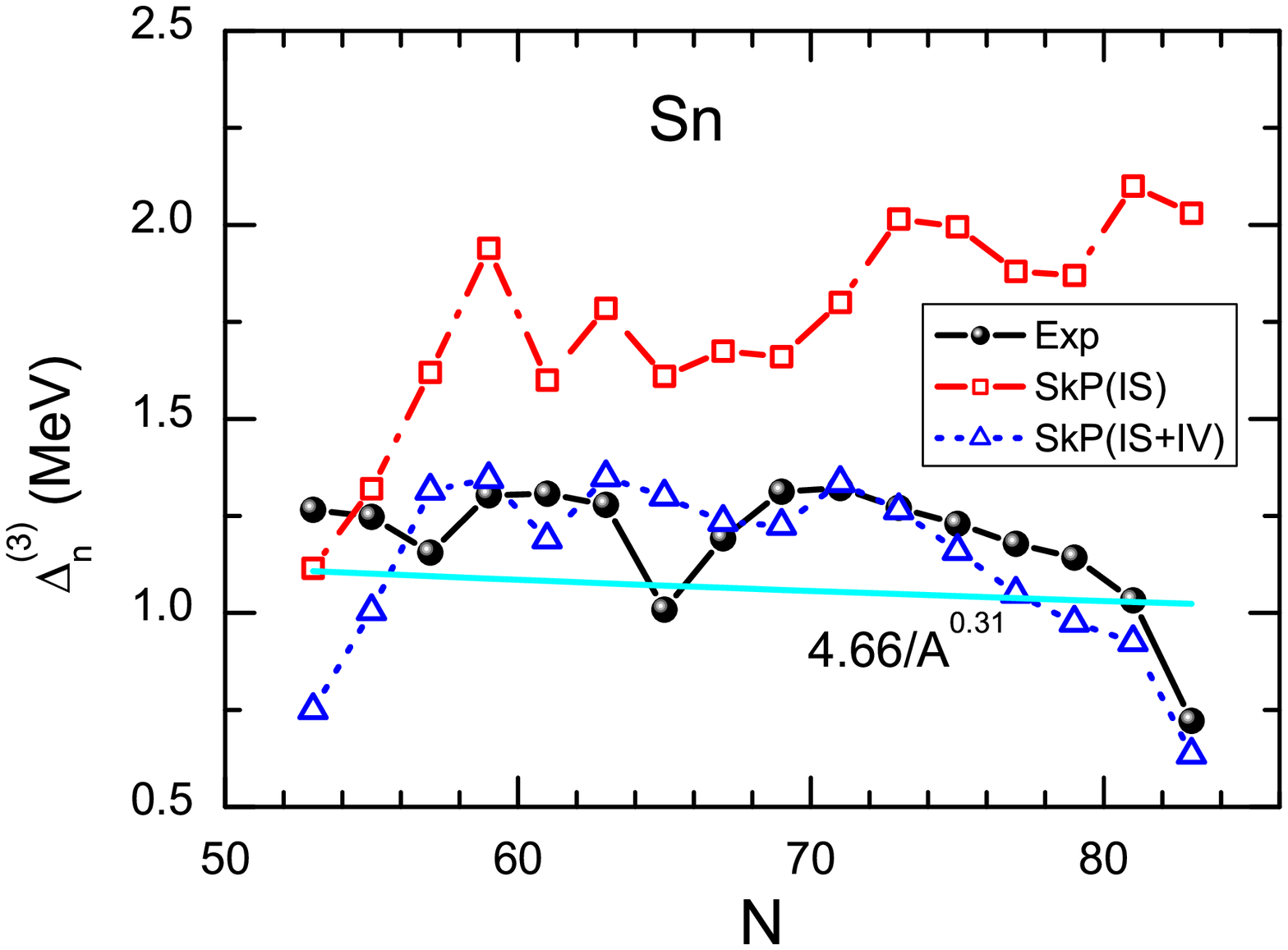}
\vspace{-0.5cm}
\includegraphics[clip,scale=0.375]{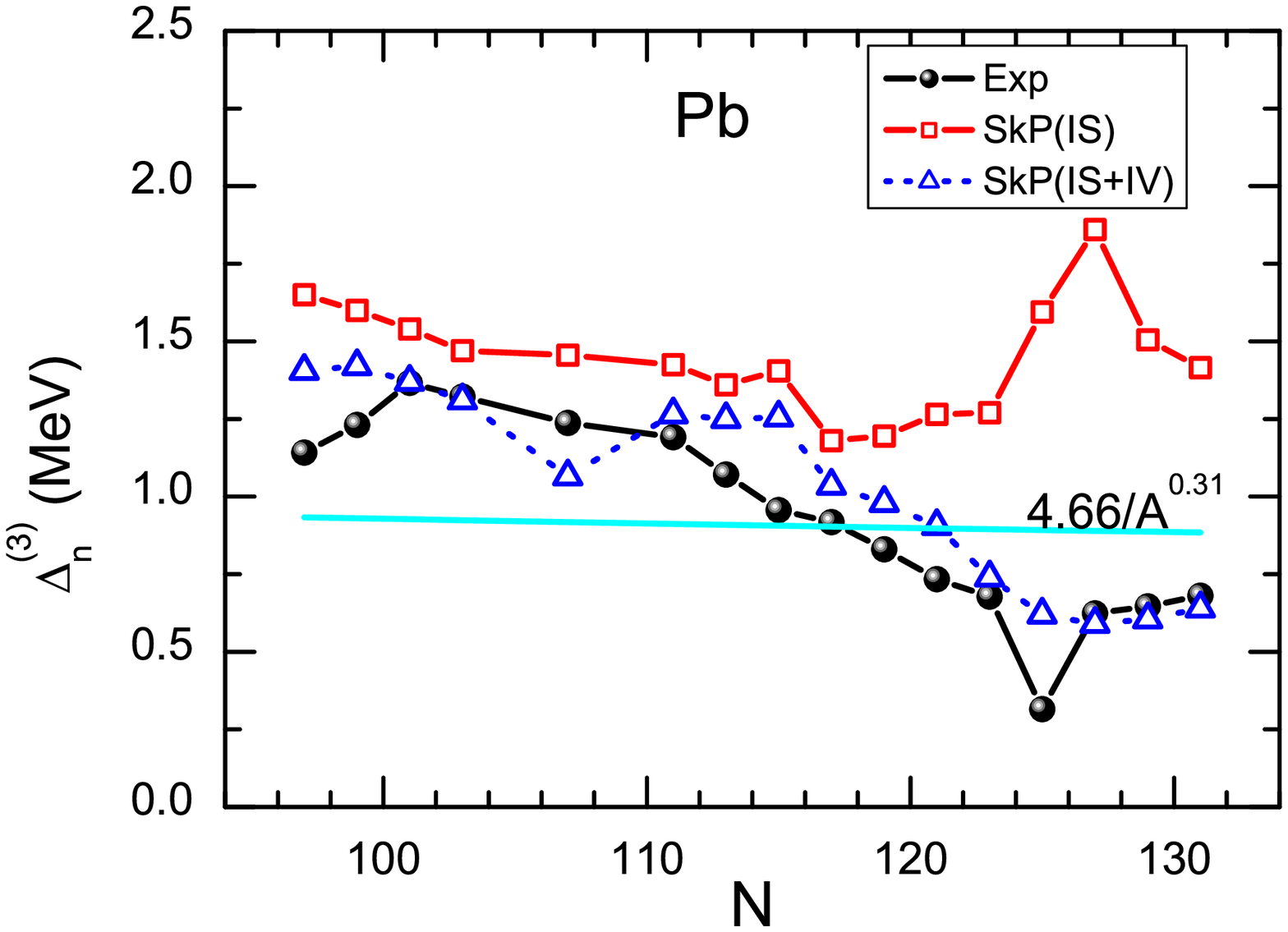}
\caption{(Color online) Odd-even mass staggering
$\Delta_\mathrm{n}^{(3)}$ calculated  by
Eq. \eqref{eq:oes} for  the semi-magic Sn and Pb isotopes.
The SkP interaction is adopted together with the IS pairing  \eqref{isoscalar} or  the IS+IV pairing \eqref{eq:g1t} in HF+BCS model.
The filling approximation is applied to the last unoccupied particle
in odd nuclei.  See the text for details.} \label{fig01}
\end{figure}

\begin{figure}[htb]
\vspace{-0.5cm}
  \includegraphics[clip,scale=0.375]{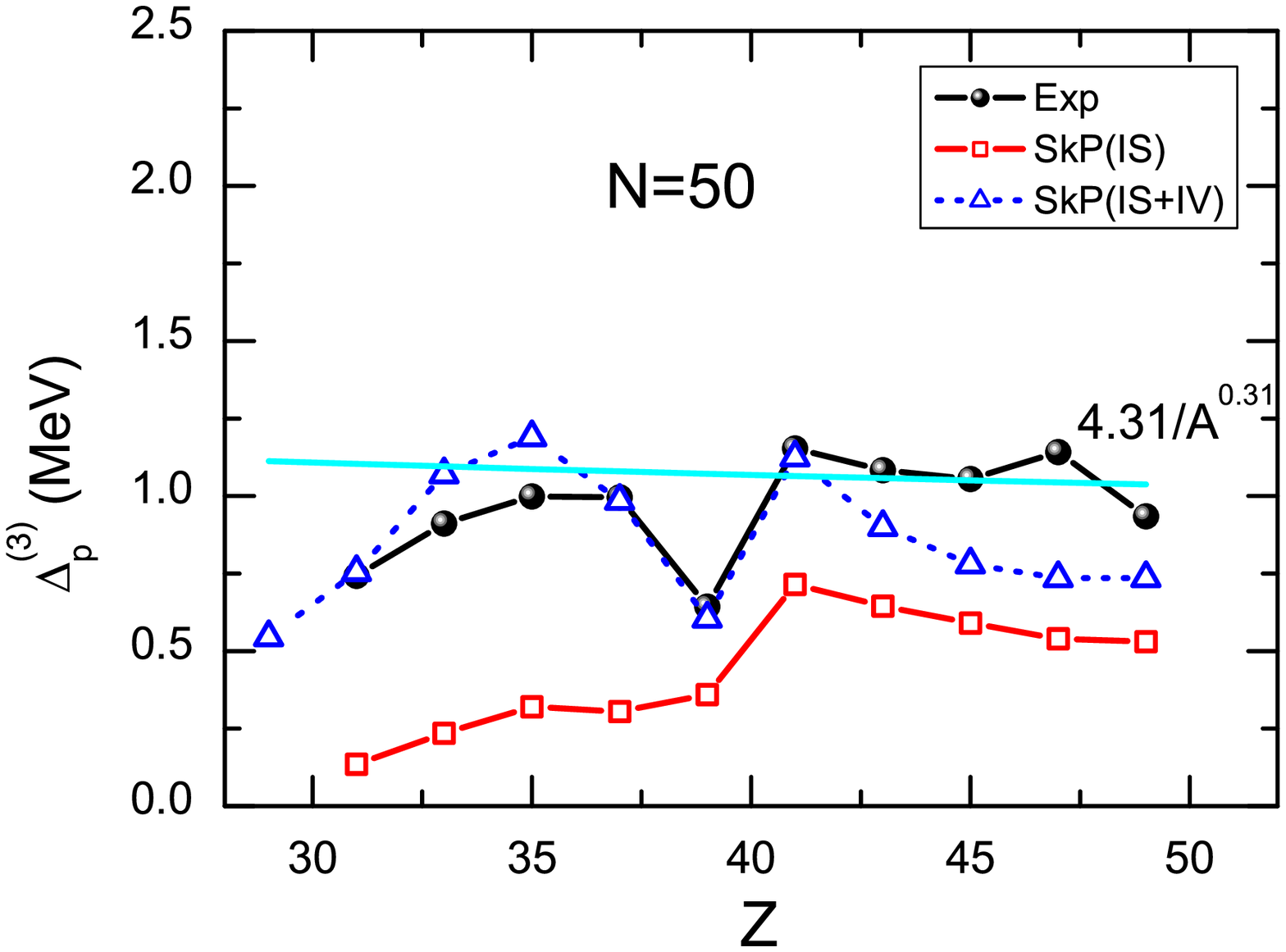}
\vspace{-0.5cm}
  \includegraphics[clip,scale=0.375]{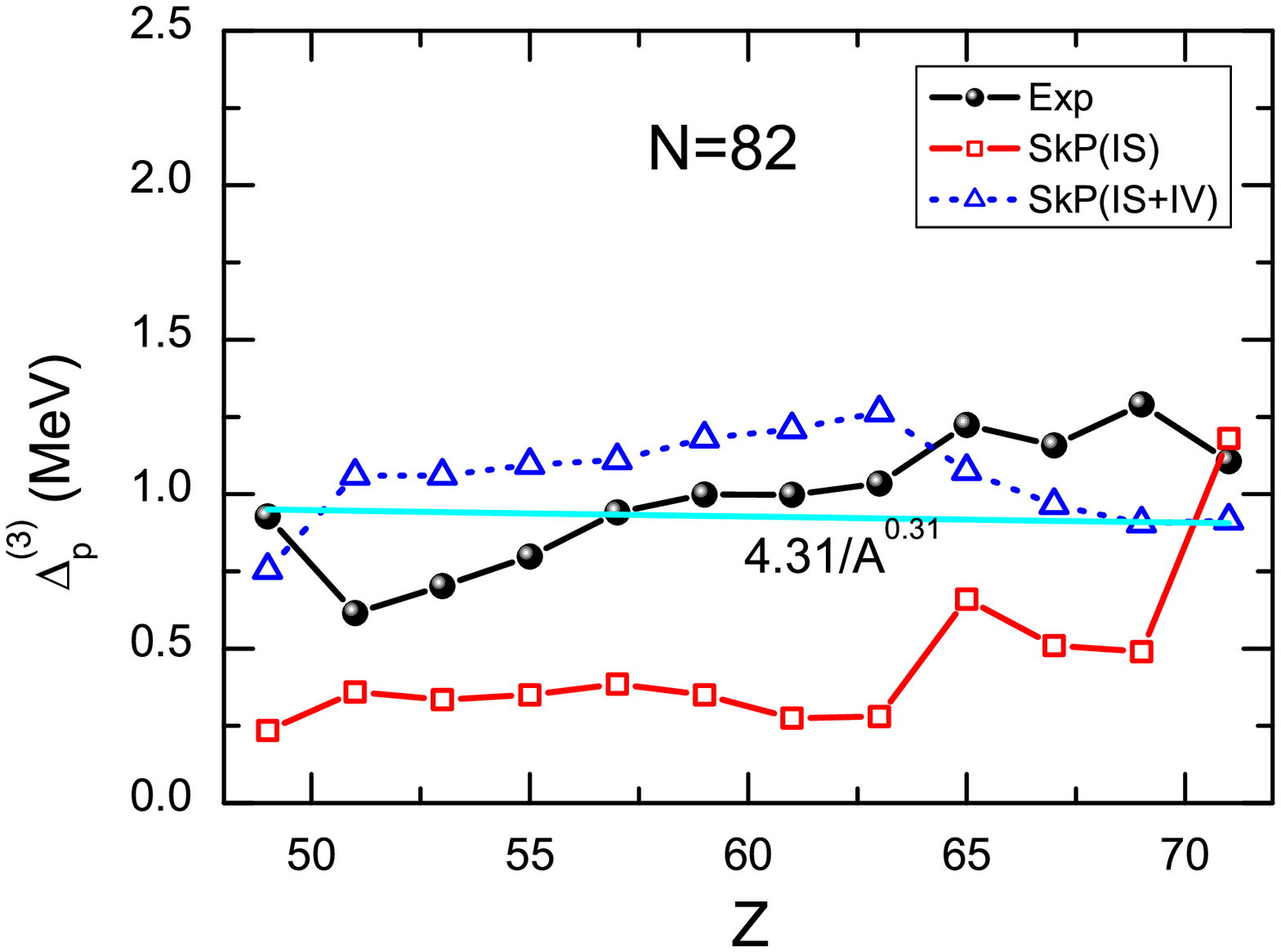}
\caption{(Color online) Odd-even mass staggering
$\Delta_\mathrm{p}^{(3)}$ calculated  with
Eq. \eqref{eq:oes} for  the N=50 and N=82 isotones.
The SkP interaction is adopted together with the IS pairing  \eqref{isoscalar} or  the IS+IV pairing \eqref{eq:g1t} in HF+BCS model. See
the caption to Fig. \ref{fig01} and the text for details.
 } \label{fig02}
\end{figure}

\begin{figure}[htb]
\vspace{-0.5cm}
\includegraphics[clip,scale=0.375]{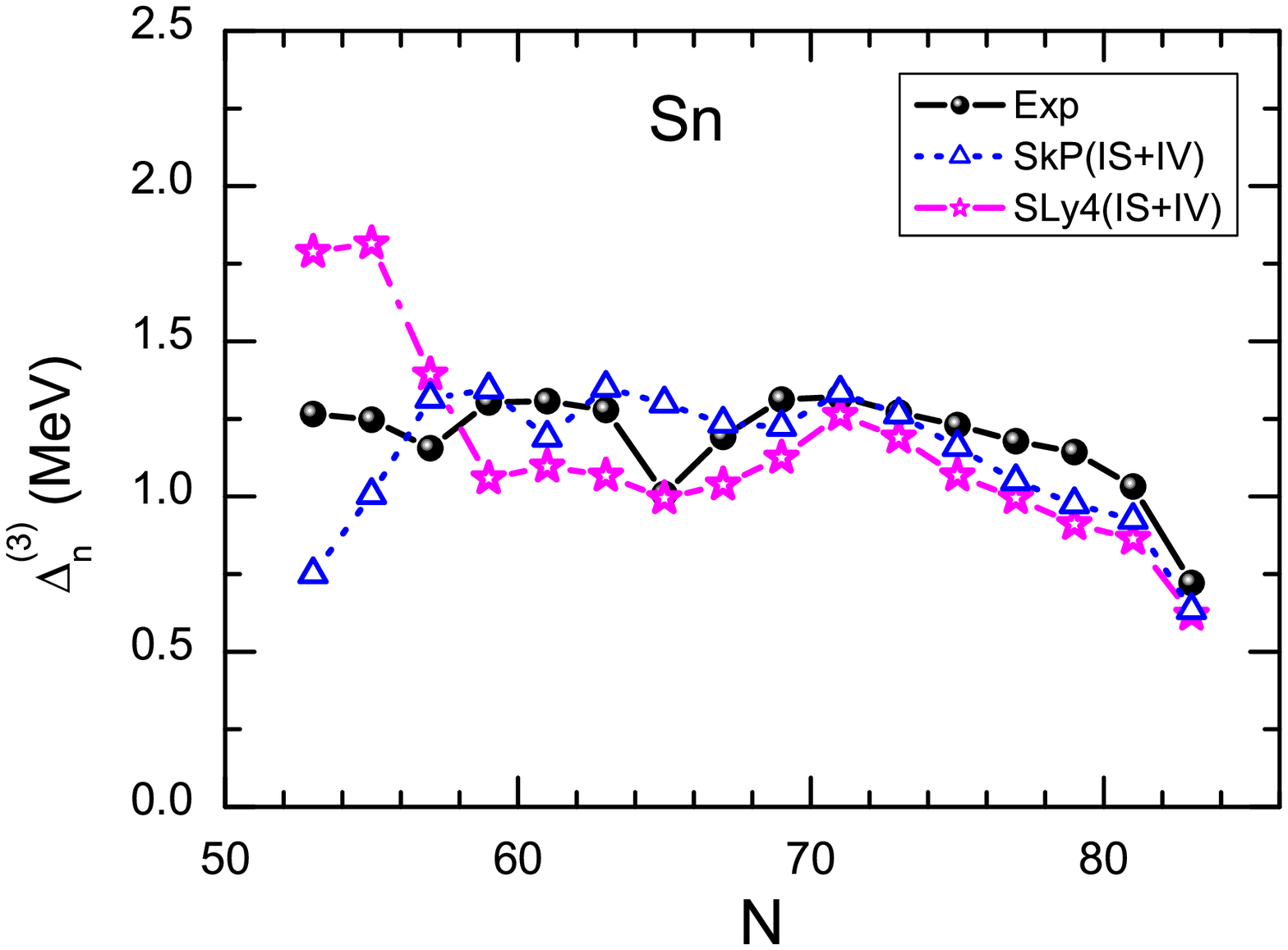}
\vspace{-0.5cm}
\includegraphics[clip,scale=0.375]{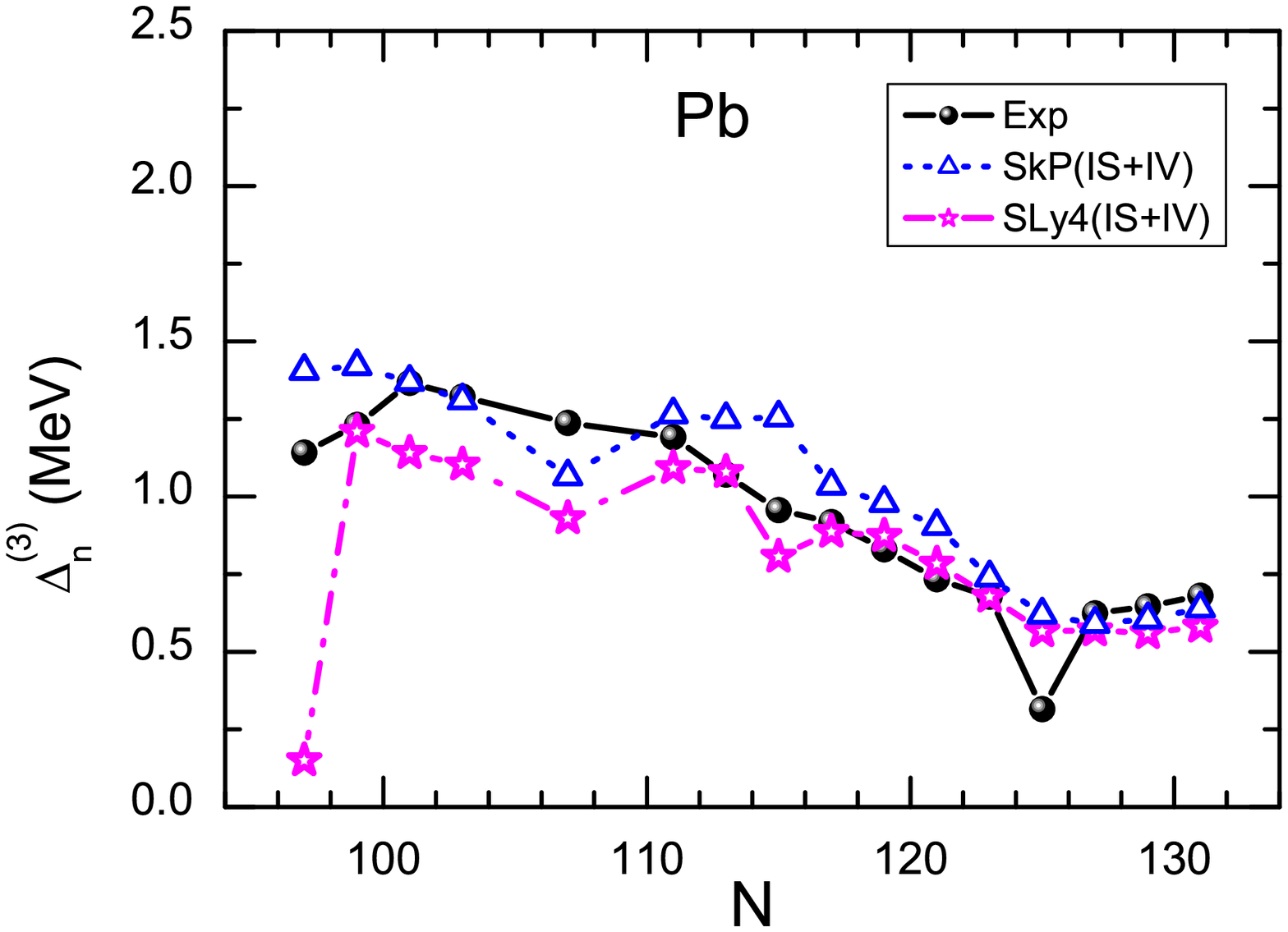}
\caption{(Color online) Odd-even mass staggering
$\Delta_\mathrm{p}^{(3)}$ calculated  with
Eq. \eqref{eq:oes} for  the N=50 and N=82 isotones.
The SLy4 interaction is adopted together with the IS pairing
  \eqref{isoscalar} or  the IS+IV pairing \eqref{eq:g1t} in HF+BCS model. See
the caption to Fig. \ref{fig01} and the text for details.} \label{fig03}
\end{figure}

The calculated results are shown in Figs. \ref{fig01}-\ref{fig03}.
    The HF+BCS results
are compared with the experimental data and also the
phenomenological parameterization 
\begin{equation}
 \bar{\Delta}=c/A^{\alpha}
\label{eq:gap-pheno}
\end{equation}
with $c=4.66(4.31)$ MeV for neutrons (protons)
and $\alpha=$0.31 which gives the rms residual of
0.25 MeV \cite{bertsch09}.

Figure \ref{fig01} shows the OES  $\Delta_n^{(3)}$ for Sn and Pb
isotopes. The calculations are performed with the SkP interaction.  The
overall agreement with the IS+IV pairing interaction \eqref{eq:g1t}
gives quite satisfactory results. Compared with the results with IS
pairing \eqref{isoscalar},  the difference is clearly seen in
neutron-rich isotopes while the difference is rather small in
neutron deficient isotopes.  The difference of the two results in
larger isospin nuclei is induced by the isospin dependence in Eq.
\eqref{eq:g1t} which weakens the pairing strength effectively in
neutron-rich nuclei. The experimental OES $\Delta_n^{(3)}$ for Sn
isotopes is rather constant  around 1.2 MeV until N=80 and decease
below 1 MeV above N=82. This trend is well reproduced  by the IS+IV
pairing. On the other hand, the calculated results increase
gradually as a function of N and reach up to 2 MeV in heavier Sn
isotopes.  This feature  certainly does not agree with the
experimental one.  The experimental  $\Delta_n^{(3)}$ for Pb
isotopes is about 1.4 MeV in neutron-deficient Pb isotopes and go
down to 0.7 MeV in neutron-rich isotopes.  This trend is again well
accounted by the IS+IV pairing while the IS pairing fails to
reproduce this trend in neutron-rich isotopes. The phenomenological
gap formula \eqref{eq:gap-pheno} gives good account of overall OES
in medium-heavy and heavy nuclei. The average values of $\Delta_n^{(3)}$
for Sn and Pb isotopes are also well reproduced by this formula, but the
isospin dependence is relatively weak compared to the experiments
and also the IS+IV results, especially in Pb isotopes.
  In Pb  isotopes, the formula gives
 0.93  MeV and 0.90 MeV for N=99 and 121 isotopes, respectively, while the
experimental values are 1.23 MeV and 0.73 MeV  for the
corresponding isotopes.

In Fig. \ref{fig02}, the calculated proton OES  $\Delta_p^{(3)}$ are
shown together  with the experimental data of N=50 and N=82
isotones.  The IS+IV pairing gives again better  agreement with the
experimental data than  the IS one.  Notice that the IS+IV pairing
strength becomes larger  effectively for smaller Z isotones because
of the isospin factor $\tau_z=-1$ for protons in the interaction
\eqref{eq:g1t}. Quantitatively,  the IS pairing gives only about the half of
 the experimental values, even less than half for N=82 isotones.
On the other hand, the IS+IV pairing provides proper amount of the OES
in both N=50 and N=82 isotones because of larger pairing strength
for protons in proton deficient isotones.
The kink at Z=39 in the Fig. \ref{fig01} for N=50 isotones is
due to the subshell  structure at  Z=40 which is also appeared in
the curve of  isospin dependent pairing (the white triangles). The
formula \eqref{eq:gap-pheno} gives the proton OES to be 1.10 and
1.04 MeV for Z=31 and Z=47 of N=50 isotones respectively, while the
experimental values are 0.74 and 1.14 MeV for Z=31 and  Z=47
isotones, respectively.
We should remind that the Coulomb interaction might play a role for
proton OES which is discarded in the present calculations.
Some renormalization of the effective pairing strength V$_0$
might be needed to study the proton OES under  the effect of the
Coulomb interaction.

In Fig. \ref{fig03}, the IS+IV pairing is tested against another
 interaction SLy4 for Sn and Pb isotopes.  The general features for SLy4 are
 quite similar to those of SkP except very neutron deficient isotopes.
 This might be due to the small energy window of EV8, but not real
  physical effect due to the different interactions.
  Thus the IS+IV pairing works well for OES
 irrespective of  the Skyrme interactions SkP and SLy4.


In summary, we studied the neutron OES of Sn and Pb isotopes and also the proton
OES of N=50 and N=82 isotones by using HF+BCS model with SkP and SLy4 
interactions together with the isospin dependence pairing (IS+IV
pairing)
 and IS pairing interactions.  The calculations are performed with the EV8 code for
even-even nuclei and also even-odd nuclei using the filling approximation.
  For the
 neutron pairing gaps,  the IS+IV  pairing strength
 decreases gradually as a function
 of the asymmetry parameter $(\rho_n(r)-\rho_p(r))/\rho(r)$.  On the other hand,
 the strength for protons
  is increasing for larger values of the asymmetry parameter
 because of the isospin factor in Eq. \eqref{eq:g1t}.
  The isotope dependence of the neutron OES $\Delta^{(3)}_n$ is well
  reproduced by the present calculations with the isospin dependent pairing
 compared with the IS pairing.   We can also see  the good agreement between
 the experimental proton OES and the calculations with the
 isospin dependent pairing for N=50 and N=82 isotones.
  
  We tested the IS+IV pairing for the Skyrme interaction SkP  
  and found almost the same quantitative agreement  as with
SLy4, i.e., the results reproduces well
 the experimental data of Sn and Pb isotopes.
Thus, we confirm the  clear manifestation  of the isospin
 dependence of the   pairing interaction in the OES 
in comparison with the experimental data 
 both for protons and neutrons.
  More comprehensive study of OES in the entire mass region is
 planed as a future work.

This work was partially supported  by the U.S. DOE
grants DE-FG02-08ER41533 and DE-FC02-07ER41457 (UNEDF, SciDAC-2),
and the JUSTIPEN/DOE grant DEFG02- 06ER41407, and the Scientific
Research Foundation of China Agriculture University under Grant No.
2007005, and by the Japanese Ministry of Education, Culture, Sports,
Science and Technology by Grant-in-Aid for Scientific Research under
the Program number C(2) 20540277. Computations were carried out on
the Athena cluster of the University of Washington.

\end{document}